\documentclass[sigconf, screen]{acmart}
\AtBeginDocument{%
  }

\setcopyright{acmlicensed}
\copyrightyear{2026}
\acmYear{2026}
\acmDOI{XXXXXXX.XXXXXXX}
\acmConference[MM '26]{Proceedings of the 34th ACM International Conference on Multimedia}{November 10--14, 2026}{Rio de Janeiro, Brazil}
\acmISBN{978-1-4503-XXXX-X/2018/06}

\usepackage{float}
\usepackage{subcaption}
\usepackage{colortbl}
\renewcommand\footnotetextcopyrightpermission[1]{} %remove copyright
\begin{document}

\title{GEAR: Reconstruction of Classical Paintings via Geometry Grounding and Appearance Restitution}

\author{Qinyu Zhang}
\affiliation{%
  \institution{Northwest University}
  \city{Xi'an}
  \state{Shaanxi}
  \country{China}
}

\author{Xinda Liu}
\authornote{Corresponding author.}
\email{liuxinda@nwu.edu.cn}
\affiliation{%
  \institution{Northwest University}
  \city{Xi'an}
  \state{Shaanxi}
  \country{China}
}

\author{Yunchen Li}
\affiliation{%
  \institution{Northwest University}
  \city{Xi'an}
  \state{Shaanxi}
  \country{China}
}

\author{Yunzhuo Liu}
\affiliation{%
  \institution{Northwest University}
  \city{Xi'an}
  \state{Shaanxi}
  \country{China}
}

\author{Chenxi Hu}
\affiliation{%
  \institution{Northwest University}
  \city{Xi'an}
  \state{Shaanxi}
  \country{China}
}

\author{Kang Li}
\affiliation{%
  \institution{Northwest University}
  \city{Xi'an}
  \state{Shaanxi}
  \country{China}
}

\author{Guohua Geng}
\affiliation{%
  \institution{Northwest University}
  \city{Xi'an}
  \state{Shaanxi}
  \country{China}
}


\begin{abstract}
Classical paintings preserve rich spatial, cultural, and historical content, making their reconstruction as explorable 3D scenes valuable for digital preservation, immersive exhibition, and cultural engagement. Yet, unlike photographs, they often depict scenes in a single-view, stylized manner, with weak perspective, lighting, and depth cues. Existing 3D reconstruction methods are largely built on natural-image priors, making it difficult to recover geometrically plausible and visually faithful 3D representations from such inputs. To address this challenge, we introduce \textbf{Classical Painting-to-3D (CP3D)}, a new task that aims to recover a 3D representation from a single classical painting while jointly ensuring geometric plausibility, appearance fidelity to the source artwork, and plausible novel-view synthesis. We further propose \textbf{GeAR}, a training-free two-stage framework for Geometry Grounding and Appearance Restitution. GeAR first converts the input painting into a geometry-grounded representation with more coherent shading and illumination cues, improving the stability of 3D Gaussian reconstruction. It then restores artwork-faithful appearance across views under spatial constraints and multi-view consistency, recovering the painterly textures and details weakened during grounding. In addition, we construct \textbf{HeriArch}, a curated benchmark of 10,160 high-resolution classical artworks for systematic evaluation of CP3D. Extensive experiments and user studies show that GeAR consistently outperforms strong baselines in geometric plausibility, appearance fidelity, and human preference. Code and dataset will be released publicly.

\end{abstract}



\begin{teaserfigure}
  \centering
  \includegraphics[width=\textwidth]{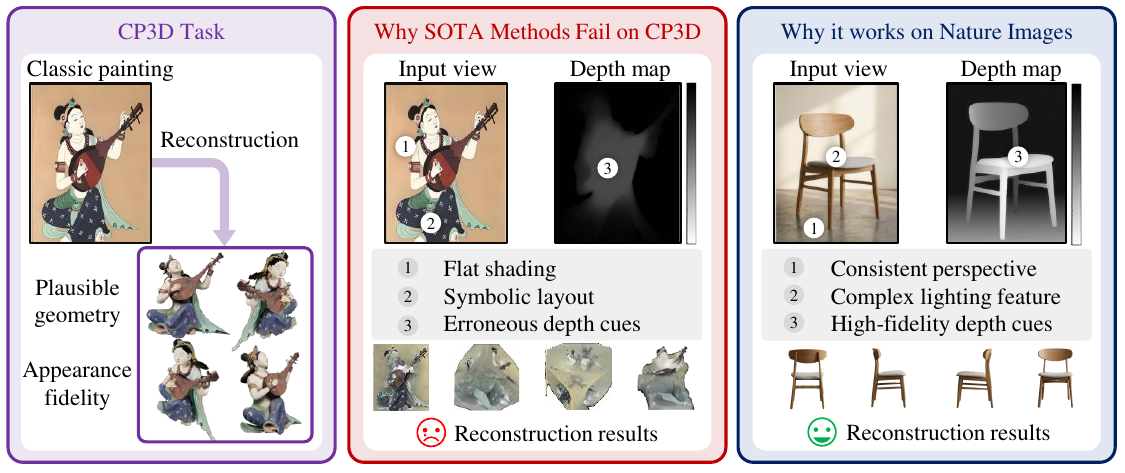}
\caption{Motivation of Classical Painting-to-3D (CP3D) and the domain gap between classical paintings and natural images. (a) CP3D aims to recover 3D structure from a classical painting while preserving appearance fidelity.
(b) Paintings provide sparse, stylized, and inconsistent geometric cues, violating the assumptions of existing reconstruction methods.
(c) Natural images provide physically consistent cues that support reliable 3D reconstruction.}
  \label{fig:motivation}
\end{teaserfigure}

\maketitle

\section{Introduction}
\label{sec:intro}
Classical paintings are enduring visual records of lost worlds, preserving not only people and events but also how past cultures organized space, ritual, and everyday life. Reconstructing them as explorable 3D scenes is therefore valuable not only for recovering the spatial experience embedded in the artwork, but also for supporting digital preservation, immersive exhibition, and new forms of cultural engagement \cite{dahn2024fate,zhao2025reviving,springstein2024visual}. Unlike photographs, however, classical paintings often rely on flattened composition, stylized perspective, and symbolic depiction rather than physically consistent scene geometry \cite{songsheng2025re,small2019circling}. Consequently, their reconstruction cannot be handled by directly applying existing single-image 3D reconstruction methods developed for natural images.

Motivated by this mismatch, we introduce a new task, Classical Painting-to-3D (\textbf{CP3D}), for inferring a plausible 3D representation from a single classical painting. CP3D requires a 3D representation that jointly satisfies geometric plausibility and appearance fidelity to the source painting, while enabling plausible novel-view synthesis (Fig.~\ref{fig:motivation} (a)). This dual requirement makes CP3D fundamentally different from standard single-image 3D reconstruction, which assumes to be a camera observation of a real, physically consistent 3D scene rather than a planar depiction shaped by artistic intent and stylized representation.

This difference makes reliable geometry recovery in CP3D particularly challenging, since the available spatial cues are often weak, stylized, and ambiguous. As shown in Fig.~\ref{fig:motivation} (b), classical paintings frequently exhibit flattened shading, symbolic layout, and misleading depth cues, making perspective, illumination, and spatial structure difficult to interpret in  physically consistent terms.  By contrast, natural images usually offer more coherent perspective, richer illumination variation, and stronger depth cues, as illustrated in Fig.~\ref{fig:motivation} (c). Existing monocular reconstruction models \cite{somraj2023vip,yang2023freenerf,szymanowicz2025flash3d,shen2025gamba} therefore struggle in the CP3D setting because their underlying assumptions, including photometric consistency, physically grounded image formation, and natural-image statistics, are frequently violated in classical paintings.

To address this tension, we propose \textbf{GeAR}, a training-free framework for \textbf{Ge}ometry Grounding and \textbf{A}ppearance \textbf{R}estitution. The key insight of GeAR is that the representation best suited for stable geometry recovery is not necessarily the one that best preserves the painterly appearance of the source artwork. In classical paintings, the visual patterns that obstruct physically grounded reconstruction often remain essential to brushwork, texture, and stylistic identity. GeAR therefore does not force a single representation to simultaneously maximize structural recoverability and painterly fidelity. Instead, it assigns these two requirements to different stages of the reconstruction pipeline. Geometry Grounding first transforms the input painting into a geometry-oriented representation with more coherent shading and illumination cues,  reducing ambiguity introduced by stylized depiction and enabling a pretrained reconstruction model to infer a more stable 3D representation. Because this process may attenuate fine textures and painterly details, GeAR then performs Appearance Restitution on the grounded 3D scene under spatial constraints and multi-view consistency, restoring the visual characteristics weakened during grounding while preserving the recovered structure. This design improves geometric stability without collapsing the reconstructed result toward a natural-image appearance, while retaining the visual identity of the original painting throughout the reconstruction process.

To support systematic study of CP3D, we construct \textbf{HeriArch}, a large-scale benchmark for classical painting-to-3D reconstruction. Existing cultural-heritage datasets \cite{xu2024comprehensive,wang2018dunhuang,li2024towards} mainly focus on 2D restoration and preservation, and are therefore not suitable for evaluating 3D reconstruction from classical planar artworks. By contrast, HeriArch contains 10,160 high-resolution images spanning murals, ukiyo-e prints, and religious paintings, covering diverse artistic styles, compositional conventions, and degrees of geometric ambiguity. This diversity makes HeriArch a suitable testbed for systematic evaluation of CP3D  across varied artistic conventions and levels of geometric ambiguity.

Our contributions can be summarized as follows:

\noindent $\bullet$ We introduce \textbf{Classical Painting-to-3D (CP3D)}, a new problem setting for inferring a plausible 3D representation from a single classical painting while jointly enforcing geometric plausibility and fidelity to the source artwork.

\noindent $\bullet$ We propose \textbf{GeAR}, a training-free framework that improves geometric recoverability by grounding classical paintings into a geometry-oriented representation, and restores painterly fidelity by recovering artwork-faithful appearance on the grounded 3D scene.

\noindent $\bullet$ We construct \textbf{HeriArch}, a 10,160-image benchmark for systematic evaluation of CP3D, and show through extensive qualitative, quantitative, and human evaluations that GeAR achieves more stable geometry and more faithful appearance than strong baselines.

\begin{figure*}[h] 
    \centering
    \includegraphics[width=\linewidth]{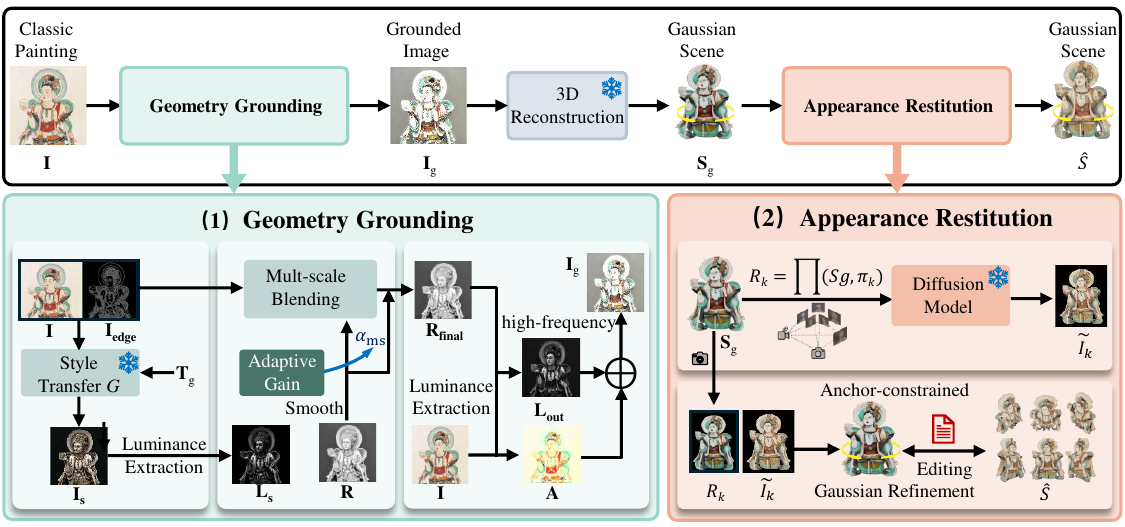}
     \caption{
    Overview of GeAR. We first perform Geometry Grounding to transform a classical painting into a geometry-oriented representation for stable 3D reconstruction, yielding a grounded Gaussian scene $\mathbf{S}_g$. We then apply Appearance Restitution, where multi-view targets guided by a source-derived prompt are used to refine the scene via anchor-constrained optimization. The final result $\mathbf{S}^*$ is both geometrically plausible and faithful to the original artwork.
    }

    \label{fig:framework}
    \vspace{-1em}
\end{figure*}

\section{Related Work}

\noindent\textbf{Neural Single-Image 3D Reconstruction.} 
Reconstructing 3D scenes from a single image is a fundamental yet ill-posed task in computer vision, owing to missing depth cues and occluded geometry \cite{huang2025spar3d,chen2024recon3d}. Existing methods often adopt either neural implicit representations \cite{mildenhall2021nerf,niemeyer2022regnerf,somraj2023vip} or point-based explicit approaches \cite{kerbl20233d,charatan2024pixelsplat,fan2024instantsplat}. 
NeRF \cite{mildenhall2021nerf} models continuous 3D scenes via MLPs but struggles under sparse-view inputs. Recent extensions introduce structural priors for regularization, such as depth smoothing \cite{niemeyer2022regnerf}, visibility priors \cite{somraj2023vip}, and frequency constraints \cite{yang2023freenerf}. 
In parallel, monocular 3D Gaussian Splatting (3DGS) is gaining traction. Splatter Image \cite{szymanowicz2024splatter} maps pixels to Gaussians in a feed-forward manner. Flash3D \cite{szymanowicz2025flash3d} predicts layered Gaussians for occlusion modeling. Triplane-based approaches \cite{shue20233d} and transformer-based pipelines \cite{shen2025gamba} further improve efficiency. 
However, these methods are largely developed for natural images and typically assume physically consistent scene geometry, perspective, and appearance cues, assumptions that may not hold for classical or heritage paintings.

\noindent\textbf{Generative 3D via Diffusion and Latent Priors.}
Diffusion models and latent priors have recently been exploited for 3D reconstruction. 
SiTH produces unseen-view textures and recovers textured meshes \cite{ho2024sith}, while Diffusion‑FOF models Fourier Occupancy Fields for joint geometry and appearance. 
VideoScene distills video diffusion into a one-shot 3D scene generator \cite{wang2025videoscene}, and DRiVE creates riggable 3D Gaussians from a single image \cite{sun2025drive}. 
Other notable methods include RenderDiffusion, which integrates rendering into the denoiser to enforce multi-view consistency \cite{anciukevivcius2023renderdiffusion}. LN3Diff, which diffuses over a compact latent 3D space for fast monocular reconstruction and generation \cite{lan2024ln3diff}, and DiffRF, which directly generates volumetric radiance fields under rendering guidance \cite{muller2023diffrf}. 
These approaches are also primarily developed for natural-image domains, and their assumptions may not transfer directly to heritage imagery with non-physical perspective and artist-mediated structure.

\noindent\textbf{Artistic and Heritage Image Datasets.}
Existing cultural heritage datasets, including MuralDH \cite{xu2024comprehensive}, DunHuang-Mural \cite{wang2018dunhuang}, and ARMCD \cite{li2024towards}, mainly focus on 2D restoration problems and therefore provide annotations such as synthetic degradations, restoration targets, or segmentation masks rather than explicit 3D structural supervision. This design makes them valuable for preservation-oriented image restoration, but less suitable for studying monocular 3D reconstruction. In addition, their samples often contain repeated content, local crops, or region-specific annotations, which weakens the global compositional and spatial cues needed for geometric reasoning.
HeriArch is constructed to fill this gap. It is a large-scale benchmark designed for monocular 3D reconstruction from heritage imagery, containing unique samples with diverse artistic styles and cultural backgrounds. To the best of our knowledge, HeriArch is the first large-scale benchmark for evaluating monocular 3D reconstruction across a broad spectrum of non-photorealistic heritage artworks. HeriArch offers a more suitable testbed for 3D-aware cultural heritage analysis thanks to its non-redundant samples, stylistic diversity, and consistent benchmark setting.

\section{Method}
We first formalize the Classical Painting-to-3D (CP3D) task and then present GeAR, a training-free framework that decomposes reconstruction into geometry grounding and appearance restitution. The key idea is to avoid forcing a single representation to simultaneously maximize geometric recoverability and painterly fidelity. Instead, GeAR first converts the input painting into a geometry-oriented representation that is more suitable for stable 3D reconstruction, and then restores artwork-faithful appearance on the recovered 3D scene under structural constraints.

\subsection{Problem Formulation}
Given a single classical painting $\mathbf{I}\in\mathbb{R}^{H\times W\times 3}$, CP3D aims to recover a 3D scene representation $\hat{\mathbf{S}}$ that supports plausible novel-view rendering while remaining faithful to the source artwork:
\begin{equation}
\hat{\mathbf{S}}=\{g_i\}_{i=1}^{N},
\end{equation}
where $N$ denotes the number of Gaussian primitives, and each primitive $g_i$ is parameterized by its center, scale, rotation, opacity, and appearance attributes. Let $\Pi(\mathbf{S},\pi)$ denote the differentiable rendering of a Gaussian scene $\mathbf{S}$ under camera pose $\pi$. 

Unlike conventional single-image 3D reconstruction, CP3D takes as input an artist-mediated planar depiction rather than a physical observation of a geometrically consistent 3D scene. Consequently, perspective, shading, and depth cues in the painting may be stylized, ambiguous, or internally inconsistent, making reliable geometry inference difficult. At the same time, the reconstructed scene should remain faithful to the painterly appearance of the source artwork. CP3D therefore requires jointly satisfying two objectives: recovering geometrically plausible scene structure from an underconstrained painting and preserving fidelity to the source artwork in the rendered appearance.

\subsection{GeAR}
Our proposed GeAR consists of two  components, as illustrated in Figure~\ref{fig:framework}: \textbf{Ge}ometry Grounding and \textbf{A}ppearance \textbf{R}estitution.  Geometry Grounding first improves geometric recoverability from classical paintings, whose structural, shading, and illumination cues are often weak, stylized, or unreliable for direct 3D reconstruction. Specifically, the input painting is mapped to a geometry-oriented representation that provides more coherent cues for pretrained monocular reconstruction, thereby producing an initial 3D scene with stabilized geometry. Appearance Restitution then restores the painterly appearance and artwork-specific details that may be weakened during geometry grounding. Specifically, the grounded 3D scene is refined with view-aligned appearance targets and constrained Gaussian refinement, thereby recovering fidelity to the source artwork while preserving the grounded geometry.

\subsubsection{Geometry Grounding}
\label{sec:geometry_grounding}

The goal of Geometry Grounding is to stabilize 3D reconstruction from an underconstrained painting by constructing a geometry-oriented intermediate representation with more coherent structural and illumination cues. This stage is not intended to preserve the final artwork appearance. Instead, it focuses on suppressing depiction-specific ambiguities that would otherwise be absorbed into the recovered geometry.

\textbf{Structure-preserving geometry translation.}
A straightforward translation from paintings to a more natural-image-like domain may alter composition or distort object boundaries, which is undesirable because subsequent 3D recovery depends strongly on spatial layout. To preserve scene layout during translation, we first extract a contour-based structural prior:
\begin{equation}
\mathbf{I}_{\text{edge}} = f_{\text{edge}}(\mathbf{I}),
\end{equation}
where $f_{\text{edge}}(\cdot)$ denotes an edge detector applied to the input painting. The resulting contour map preserves object boundaries and coarse scene organization, which serve as structural guidance during translation. Conditioned on this structural prior, we then generate a geometry-oriented image:
\begin{equation}
\mathbf{I}_{s} = G(\mathbf{I}_{\text{edge}}, \mathbf{I}, T_g),
\end{equation}
where $G(\cdot)$ denotes a text-guided image translation model, and $T_g$ is a prompt that encourages physically coherent shape, shading, and illumination cues. This translation is not intended to beautify the painting, but to suppress depiction-specific ambiguity while retaining the original compositional structure. The translated image therefore provides a more suitable input for downstream 3D reconstruction.

\textbf{Illumination grounding.}
Building on the structure-preserving translated image, we further improve geometric recoverability by grounding its illumination cues. Because monocular 3D reconstruction depends strongly on shading and luminance structure, this step regularizes the translated image toward more spatially coherent illumination, yielding a more reliable input for downstream reconstruction.

Let $L_s$ denote the luminance of the translated image $\mathbf{I}_{s}$ after linearization to the radiance domain. We first compute a log-ratio illumination field:
\begin{equation}
R = \log(L_s+\varepsilon)-\log(\mathrm{gmean}(L_s+\varepsilon)),
\end{equation}
where $\varepsilon > 0$ avoids numerical instability, and $\mathrm{gmean}(\cdot)$ denotes the geometric mean over all pixels.
The resulting ratio field $R$ captures relative illumination variation while reducing sensitivity to the global intensity scale. We then apply local mean filtering to obtain a smoothed ratio field that enforces spatial consistency in illumination variation. For each pixel $p$, the smoothed ratio field is computed as:
\begin{equation}
   R_{\text{smooth}}(p)=\frac{1}{|\Omega(p)|}\sum_{q\in\Omega(p)} R(q), 
\end{equation}
where $p$ indexes a pixel location, $\Omega(p)$ denotes the local filter window centered at $p$, $q$ indexes pixels inside that window, and $|\Omega(p)|$ is the number of pixels in $\Omega(p)$. To balance global illumination consistency against local structural detail, we further refine $R_{\text{smooth}}$ in a multi-scale, boundary-aware manner. Specifically, we construct an illumination pyramid to capture global trends at coarse scales and local shading variations at fine scales, while deriving inverse boundary weights from the edge map to suppress halo artifacts near structural boundaries.  The resulting multi-scale estimate is denoted by $R_{\text{ms}}$.
A fixed fusion weight cannot accommodate the large variation in illumination contrast across paintings: it tends to over-regularize some cases while under-regularizing others. We therefore adapt the fusion strength on a per-image basis.

\textbf{Adaptive Gain Control.}
To this end, we introduce an adaptive gain control mechanism that regulates the fusion strength according to the global contrast of the multi-scale illumination estimate.
Specifically, we measure the global contrast of the multi-scale ratio field using its standard deviation:
\begin{equation}
    \sigma_R = \mathrm{std}(R_{\text{ms}}),
\end{equation}
which quantifies the overall contrast of the multi-scale illumination estimate.
We then define an adaptive fusion weight as:
\begin{equation}
\tilde{\alpha}_{\text{ms}} = \alpha_{\text{base}} \cdot \frac{\alpha_{\text{target}}}{\sigma_R + \varepsilon},
\end{equation}
where $\alpha_{\text{base}}$ is the base fusion strength, $\alpha_{\text{target}}$ denotes the target contrast level, and $\varepsilon>0$ is a small constant for numerical stability.
To ensure stable adaptation, we constrain the fusion weight to lie within a bounded range:
\begin{equation}
\alpha_{\text{ms}} = \operatorname{clip}(\tilde{\alpha}_{\text{ms}}, \alpha_{\min}, \alpha_{\max}),
\end{equation}
where $\operatorname{clip}(\cdot,\cdot,\cdot)$ restricts the fusion weight to $[\alpha_{\min}, \alpha_{\max}]$ for numerical stability and perceptual coherence. This form inversely scales the fusion weight with the contrast of $R_{\text{ms}}$, so that images with stronger illumination variation receive weaker smoothing, while flatter cases receive stronger regularization.
The final illumination ratio is then obtained as
\begin{equation}
R_{\text{final}}=\alpha_{\text{ms}}R_{\text{ms}}+(1-\alpha_{\text{ms}})R_{\text{smooth}}.
\end{equation}
Notably, $\alpha_{ms}$ is not confined to $[0, 1]$. When $\alpha_{ms} > 1$, 
Eq.~(9) performs controlled extrapolation instead of convex blending,  which empirically amplifies informative illumination in flat paintings. 
\textbf{Illumination-consistent reconstruction.} 
We then reconstruct a geometry-grounded image in the linear radiance domain by combining the grounded illumination with reflectance estimated from the original painting. Let $L$ denote the luminance of the original painting after linearization. We recover the grounded illumination map by injecting the refined illumination ratio in the log domain:
\begin{equation}
    L_{\text{out}} = \exp\!\big(\log L + \beta\, R_{\text{final}}\big),
\end{equation}
where $\beta$ controls the strength of illumination grounding. We estimate the reflectance map $A$ from the original painting by dividing its linearized RGB values by the original luminance:
\begin{equation}
   A = \frac{\mathrm{Linearize}(\mathbf{I})}{L + \varepsilon},
\end{equation}
where $\varepsilon > 0$ ensures numerical stability. The grounded linear image is then reconstructed by combining the preserved reflectance with the grounded illumination:
\begin{equation}
   \mathbf{I}_{g,\text{linear}} = A \odot L_{\text{out}}.
\end{equation}
To recover fine brushwork and local texture, we extract high-frequency components from the original artwork and add them back:
\begin{equation}
   \mathbf{I}_g = \mathbf{I}_{g,\text{linear}} + \mathrm{HF}(\mathrm{Linearize}(\mathbf{I})),
\end{equation}
where $\mathrm{HF}(\cdot)$ denotes a high-frequency extraction operator. The resulting grounded image $\mathbf{I}_g$ is finally fed into a pretrained monocular 3D reconstruction network~\cite{xiang2025structured} to obtain a grounded Gaussian scene $\mathbf{S}_g$. In practice, this grounded input yields more stable geometry than reconstructing directly from the original painting.
\subsubsection{Appearance Restitution}
\label{sec:appearance_restitution}
Although Geometry Grounding improves structural stability, it does not fully preserve the appearance of the original artwork. In particular, the grounding process tends to suppress high-frequency visual patterns that may be geometrically unreliable but remain essential to painterly appearance. Naively restoring each view independently in 2D is insufficient, because the resulting edits often introduce severe cross-view inconsistency when used as supervision for 3D refinement. To resolve this issue, we restore appearance directly on the grounded 3D scene through multi-view refinement under geometric constraints.

\textbf{Multi-view appearance target generation.}
We first render the grounded scene $\mathbf{S}_g$ from a set of sampled viewpoints $\{\pi_k\}_{k=1}^{n}$:
\begin{equation}
\mathbf{R}_k = \Pi(\mathbf{S}_g,\pi_k), \qquad k=1,\dots,n.
\end{equation}
These rendered views provide geometrically consistent observations of the grounded scene under known camera poses. We then edit each rendered view with an appearance-oriented diffusion model \cite{batifol2025flux} to recover artwork-faithful texture and style:
\begin{equation}
\tilde{\mathbf{I}}_k = D(\mathbf{R}_k, T_a),
\end{equation}
where $D(\cdot)$ denotes the diffusion-based appearance editing model, and $T_a$ is an appearance prompt derived from the source painting to encourage preservation of its artwork-specific texture, style, and semantic content. Because each target $\tilde{\mathbf{I}}_k$ is generated from a rendered view $\mathbf{R}_k$ under a known camera pose, it preserves the coarse layout and viewpoint of the current 3D representation. These restored views therefore serve as approximately view-aligned appearance targets for subsequent 3D refinement.

\textbf{Geometry-preserving appearance refinement.}
We then leverage GaussianEditor \cite{chen2024gaussianeditor} to refine the Gaussian representation so that its rendered views match the appearance-restored targets while preserving the grounded geometry. Formally, starting from the grounded initialization $\mathbf{S}_g$, we obtain the final representation $\hat{\mathbf{S}}$ by optimizing:
\begin{equation}
\hat{\mathbf{S}} =
\arg\min_{\mathbf{S}}
\sum_{k=1}^{n}
\mathcal{L}_{\text{edit}}\big(\Pi(\mathbf{S},\pi_k), \tilde{\mathbf{I}}_k\big)
+ \lambda_{\text{anc}}\mathcal{L}_{\text{anchor}}(\mathbf{S}, \mathbf{S}_g),
\end{equation}
where $\mathcal{L}_{\text{edit}}$ encourages the rendered appearance of $\mathbf{S}$ to match the restored view targets, while $\mathcal{L}_{\text{anchor}}$ regularizes the optimization toward the grounded initialization $\mathbf{S}_g$ to suppress geometric drift. In practice, the anchor term suppresses large geometric drift and keeps the refinement focused on texture, color, and painterly detail rather than structural change.

\section{The HeriArch Dataset}
The HeriArch dataset is a comprehensive collection of 10,160 unique images of classical hand-painted artworks, spanning from the 3rd to the 20th century and encompassing over 17 centuries of artistic evolution. This dataset encompasses a broad spectrum of artistic styles and traditions, offering an invaluable resource for research on style consistency and geometric structure recovery in cultural heritage artworks. The dataset is organized into several key subsets, each representing a distinct artistic tradition:
(1) \textcolor{blue}{Mural}: 2,361 images, including 1,591 from the DhMural dataset~\cite{muralnet2022} and additional murals from Dunhuang, China.  
(2) \textcolor{blue}{Court Ladies Paintings}: 775 depictions of historical court life from the Tang to Qing dynasties.  
(3) \textcolor{blue}{Tibetan Thangka}: 1,783 images spanning 13 centuries, illustrating the evolution of Tibetan Buddhist art.  
(4) \textcolor{blue}{Indian Art}: 709 artworks from the 18th–20th centuries, sourced from the Smithsonian Institution, showcasing three centuries of Indian artistic development.  
(5) \textcolor{blue}{Ukiyo-e}: 4,241 Japanese woodblock prints from the Edo period, documenting the transition from monochrome \textit{sumi-e} to multicolor \textit{nishiki-e}.  
(6) \textcolor{blue}{Persian Miniature}: 291 classical Persian manuscripts and illuminated paintings from the Safavid and Timurid eras. \textbf{Refer to Appendix Section 2 for details.}

\begin{table*}[h]
\setlength{\tabcolsep}{1.35mm}{
\centering
\begin{tabular}{l!{\vrule width 0.6pt}c!{\vrule width 0.6pt}cccccccc}  
\hline
\multicolumn{1}{l!{\vrule width 0.6pt}}{Method} 
& \multicolumn{1}{c!{\vrule width 0.6pt}}{VEND $\uparrow$} 
& \multicolumn{1}{c}{DDI$_{15^\circ}$ $\uparrow$}  
& \multicolumn{1}{c}{DDI$_{30^\circ}$ $\uparrow$}  
& \multicolumn{1}{c}{DDI$_{45^\circ}$ $\uparrow$} 
& \multicolumn{1}{c}{DDI$_{60^\circ}$ $\uparrow$}  
& \multicolumn{1}{c}{DDI$_{75^\circ}$ $\uparrow$}  
& \multicolumn{1}{c}{DDI$_{90^\circ}$ $\uparrow$} 
& \multicolumn{1}{c}{DDI$_{avg}$ $\uparrow$} \\ \hline
SplatterImage \cite{szymanowicz2024splatter} & 66.65 & 4.11 & 4.03 & 3.96 & 3.82 & 3.90 & 3.18 & 3.83 \\
LGM \cite{tang2024lgm} & 73.84 & 5.67 & 4.98 & 4.16 & 4.23 & 4.47 & 4.56 & 4.68 \\
TRELLIS \cite{xiang2025structured} & 77.64 & 6.35 & 6.30 & 6.58 & 6.96 & 8.12 & 9.82 & 7.37 \\ \hline
\rowcolor{blue!5}GeAR (Ours) & \textbf{85.29} & \textbf{10.66} & \textbf{10.56} & \textbf{10.62} & \textbf{10.87} & \textbf{11.29} & \textbf{12.29} & \textbf{11.05} \\ \hline
\end{tabular}
\caption{Quantitative comparison on final reconstructions. We report geometry-oriented metrics on the final outputs to evaluate whether the recovered geometry remains plausible after the full two-stage pipeline.}
\label{sota-search}}
\vspace{-1em}
\end{table*}

\section{Experiments}\label{Experiments}
\subsection{Experimental Setup}
\noindent\textbf{Evaluation Protocol and Metrics.}
Since ground-truth novel views are unavailable for classical paintings, CP3D cannot be adequately evaluated with a single family of metrics. We therefore evaluate the final reconstructions from two complementary perspectives: \textit{geometric plausibility} and \textit{artwork fidelity}. For geometric plausibility, we report two geometry-oriented metrics, \textit{Volumetric Extent and Normal Directional Diversity} (VEND) and \textit{Depth Detail Integrity} (DDI), computed on the final outputs to assess whether the recovered geometry remains plausible after the full two-stage pipeline. For artwork fidelity, we conduct a texture-focused MLLM evaluation and a human study on the final reconstructions to assess how well they preserve the painterly appearance of the source artwork and their overall perceptual preference. (1) VEND measures both the volumetric extent of the reconstructed shape and the directional diversity of surface normals, thereby reflecting global structural plausibility and local geometric variation. Higher VEND scores indicate reconstructions with less flattening and richer surface-orientation diversity. (2) DDI evaluates the integration of local surface detail and global volumetric structure by combining high-frequency geometric variation with shape anisotropy. Higher DDI values indicate that the reconstruction preserves finer depth variations while maintaining stable three-dimensional structure across viewpoints. (3) Inspired by \cite{li2025generation}, we conduct a texture-focused MLLM evaluation on the final reconstructions to assess how well they preserve the texture and painterly appearance of the source painting.
(4) We additionally conduct a human study on the final reconstructions to assess overall perceptual quality and human preference. The study involves over 300 participants, and each sample is evaluated at least 30 times in total under randomized presentation.
\textbf{Detailed definitions of VEND, DDI, and the full evaluation protocols are provided in the appendix.}

\noindent\textbf{Implementation details.}
Unless otherwise specified, we use the same settings across all experiments. For 3D reconstruction, we use TRELLIS as the pretrained monocular reconstruction backbone. For metric computation, each reconstructed scene is rendered from six offset viewpoints at $15^\circ$, $30^\circ$, $45^\circ$, $60^\circ$, $75^\circ$, and $90^\circ$ relative to the input view. In Geometry Grounding, we set $\alpha_{\text{base}} = 0.8$, $\alpha_{\text{target}} = 1.0$, $\alpha_{\text{min}} = 0.4$, and $\alpha_{\text{max}} = 1.5$. In Appearance Restitution, we render $n=7$ viewpoints from the grounded Gaussian scene to construct multi-view appearance targets. The rendered views of each painting are edited using the same source-derived appearance prompt $T_a$, and the edited views are used to supervise Gaussian refinement with GaussianEditor. Detailed hyperparameter settings, including the anchor weight $\lambda_{\text{anc}}$, the number of refinement iterations, and other implementation details, are provided in the appendix.

 \noindent\textbf{Comparison with SOTA methods.}
 Since CP3D is a new setting and no prior method is designed specifically for monocular 3D reconstruction of classical hand-painted artworks, we compare GeAR with three strong monocular 3D reconstruction baselines based on 3D Gaussian representations: SplatterImage \cite{szymanowicz2024splatter}, LGM \cite{tang2024lgm}, and TRELLIS \cite{xiang2025structured}. All baselines are evaluated using their official or publicly released inference pipelines on the same test set and input resolution for fair comparison.

\subsection{Quantitative Evaluation of Geometric Plausibility }
GeAR achieves the best geometry-oriented performance on the final reconstructions, as shown in Table~\ref{sota-search}. Compared with TRELLIS, GeAR improves VEND from 77.64 to 85.29 and increases the average DDI from 7.37 to 11.05, with consistent gains across all viewpoint offsets. These results indicate that GeAR produces final reconstructions with stronger volumetric plausibility and richer depth variation under novel views. The improvements also suggest that the geometric gains introduced by Geometry Grounding are preserved after Appearance Restitution, rather than being degraded by the final refinement stage. Overall, the results show that GeAR achieves more plausible final geometry than strong monocular 3D reconstruction baselines in the CP3D setting.

\begin{table}[!ht]
\setlength{\tabcolsep}{3.1mm}{
\centering
\begin{tabular}{l|c c cl}
\hline
\multicolumn{1}{l|}{Method} 
& \multicolumn{1}{c}{TF$\uparrow$} 
& \multicolumn{1}{c}{GA$\uparrow$} 
 
& \multicolumn{1}{c}{VC$\uparrow$} & AC$\uparrow$\\ \hline
SplatterImage \cite{szymanowicz2024splatter}& 6.12& 3.32& 3.92&4.36\\
LGM \cite{tang2024lgm}& 4.35& 2.79& 4.25&4.11\\
TRELLIS \cite{xiang2025structured}& 5.92& 6.54& 6.55&5.52\\ 
GeAR w/o AR &7.97& 7.43& 7.14&7.29\\
GeAR w/ AR & \textbf{8.71}& \textbf{8.03}& \textbf{7.95}&\textbf{7.97}\\\hline
\end{tabular}
%\vspace{-0.5em}
\caption{The results of different methods evaluated by MLLMs on four appearance dimensions. AR denotes Appearance Restitution.}
\label{LLM}}
\vspace{-2.5em}
\end{table}

\begin{figure}[t] 
    \centering
    \includegraphics[width=0.8\linewidth]{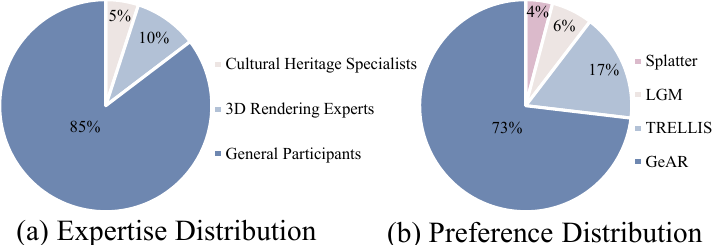}
    \caption{(a) Distribution of participants across three expertise groups; (b) Distribution of participant preferences among the evaluated methods.}
    \label{fig:userstudy}
    \vspace{-1em}
\end{figure}

\subsection{MLLM-Based Evaluation of Appearance Fidelity}
Geometric metrics alone are insufficient to assess the perceptual quality of the final reconstruction. Table~\ref{LLM} therefore reports an additional evaluation of the final outputs using a blind MLLM judge following the LLM-as-a-judge paradigm~\cite{li2025generation}, with randomized method order and four evaluation dimensions: Texture Fidelity (TF), Geometric Accuracy (GA), View Consistency (VC), and Aesthetic Consistency (AC). GeAR with Appearance Restitution achieves the best performance on all four dimensions. The gain in TF indicates that Appearance Restitution restores artwork-specific textures and painterly details weakened during Geometry Grounding, while the improvements in GA and VC suggest that the restored appearance remains compatible with the grounded geometry. The higher AC further indicates stronger overall perceptual coherence. Together, these results show that GeAR achieves a better overall balance between geometric plausibility and artwork fidelity in the final reconstruction.

\begin{figure*}[h] 
    \centering
    \includegraphics[width=0.85\linewidth]{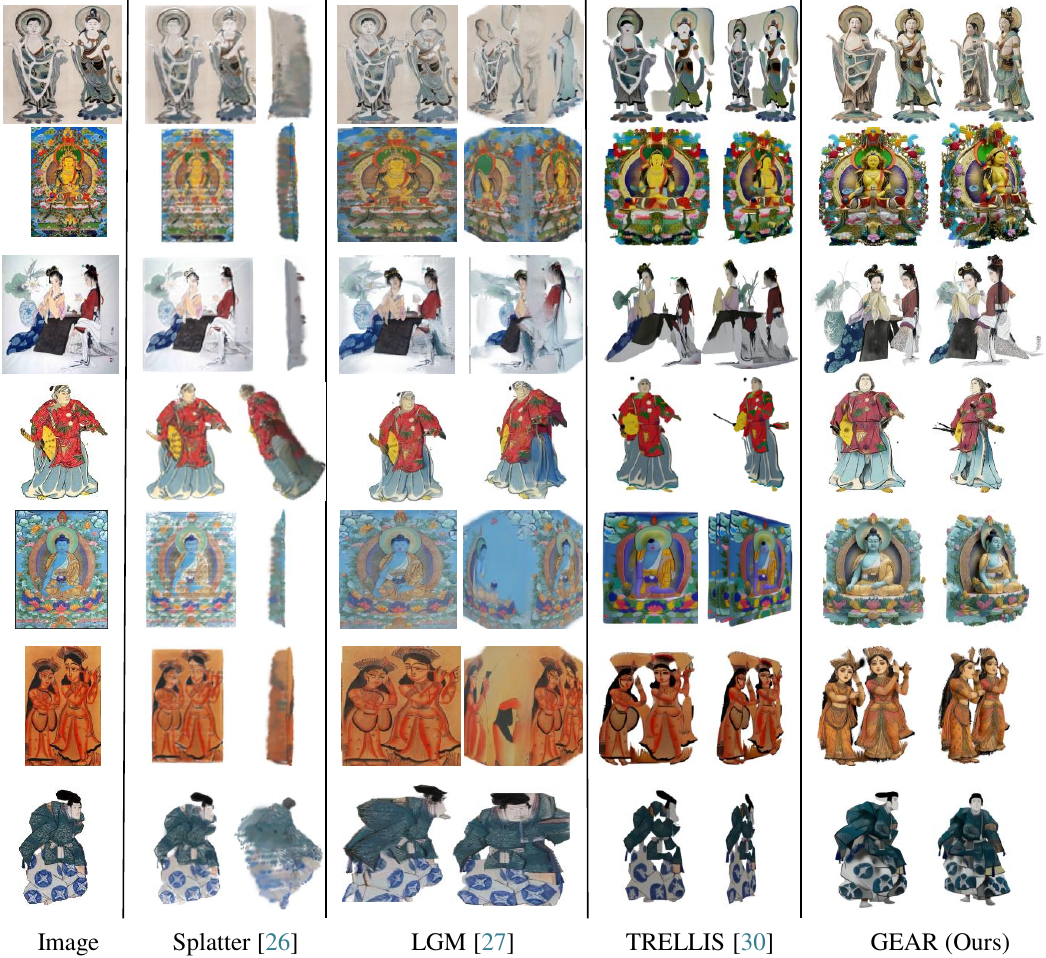}
    \caption{ Comparisons of generated 3D Gaussians for image-to-3D. Including Mural, Indian Art, Persian Miniature, Court Ladies Paintings, Tibetan Thangka, and Ukiyo-e. Our method generates Gaussian splatting with better visual quality on various challenging hand-painted artworks. Additional visual results are available in Appendix.}
    \vspace{-0.5em}
    \label{fig:visual}
    
\end{figure*}
\subsection{User Study}
We conducted a comprehensive user study involving over 300 participants, including 29 experts in 3D rendering and 15 specialists in cultural heritage. The goal was to compare different monocular 3D reconstruction methods for hand-painted artworks based on user preference. Specifically, we randomly sampled 300 classic hand-painted artworks from the HeriArch Dataset and generated corresponding monocular 3D reconstructions using various methods. The reconstructed results were randomly shuffled and presented to participants, who were asked to evaluate them from three perspectives: perceived depth realism, aesthetic quality, and overall preference based on personal intuition. As shown in Figure~\ref{fig:userstudy}, GeAR received overwhelmingly positive feedback, being strongly preferred by participants for its significant improvement in reconstruction quality.

\subsection{Qualitative Results}
We compare GeAR with recent monocular image-to-3D Gaussian reconstruction methods. Figure~\ref{fig:visual} presents rendered results from the reconstructed 3D models under different viewpoints. GeAR achieves better preservation of artistic details such as figure posture, fabric folds, and object structure. SplatterImage and TRELLIS lose substantial detail, particularly in complex regions such as faces and clothing, while LGM recovers some structure but remains blurry and distorted in high-frequency areas. In contrast, GeAR produces more accurate geometry, preserves local details more effectively, and maintains better global shape consistency. It also recovers lighting variation and material appearance more faithfully, especially in shadow transitions, fabric reflections, and object glossiness. Overall, GeAR produces more stable and visually faithful reconstructions than competing methods.

\subsection{Ablation Study}\label{ablation}
In this section, we perform ablation studies to examine the contribution of key components in GeAR. Unless otherwise specified, all results are reported on the final reconstructions. For analyses related to Geometry Grounding, we additionally report NIQE on the grounded intermediate as a supplementary diagnostic of natural-image compatibility rather than a holistic metric of final CP3D quality, since Appearance Restitution intentionally restores painting-specific textures and stylistic patterns that may deviate from natural-image statistics.
\begin{figure}[h] 
    \centering
    \includegraphics[width=\linewidth]{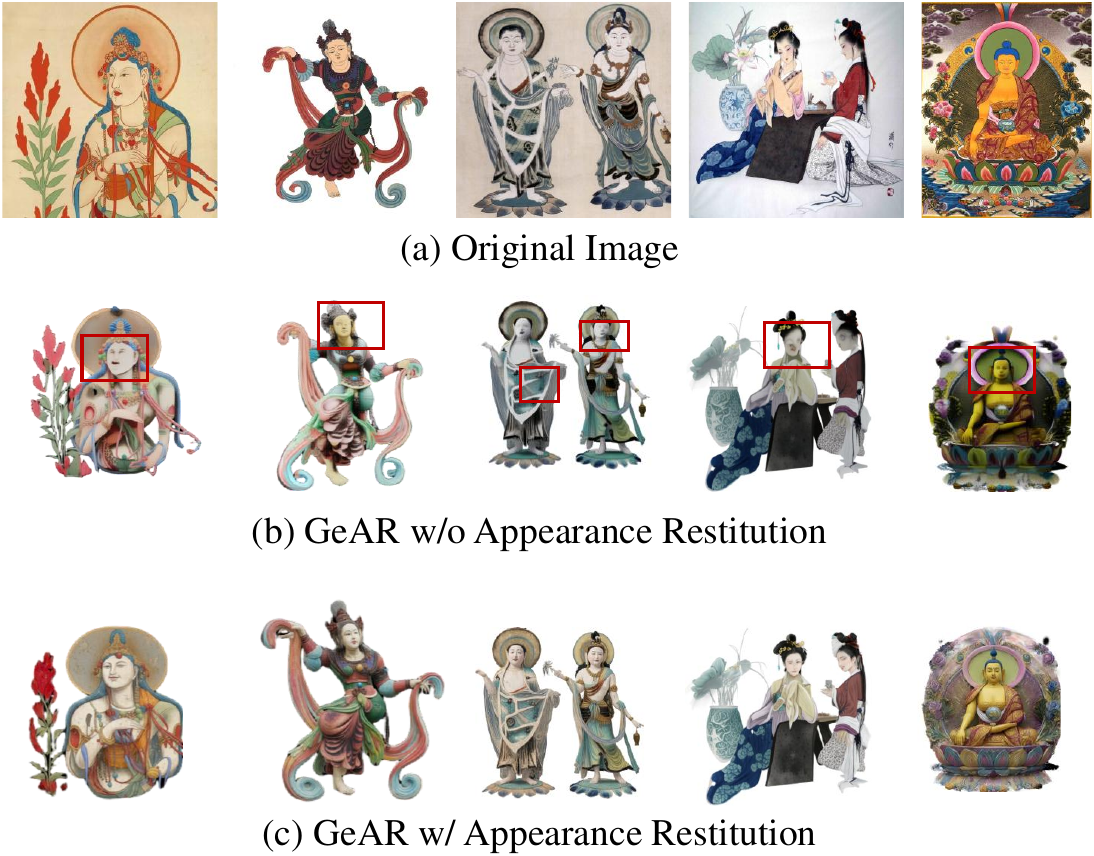}
     \vspace{-0.5em}
    \caption{ Qualitative Comparison of Reconstruction Results W/ and W/o Appearance Restitution}
  \vspace{-0.5em}
    \label{fig:v-a}
    
\end{figure}

\begin{table}[ht]
\setlength{\tabcolsep}{2.85mm}{
\centering
\begin{tabular}{l|c c c}
\hline
\multicolumn{1}{l|}{Method} 
& \multicolumn{1}{c}{VEND $\uparrow$} 
& \multicolumn{1}{c}{DDI$_{avg}$ $\uparrow$} 
& \multicolumn{1}{c}{NIQE$_{avg}$ $\downarrow$} \\ \hline
\textit{GeAR w/o AR}             & 85.84 & 6.68 & \textbf{4.75} \\
\textit{GeAR w/ AR}         & \textbf{87.61} & \textbf{10.24} &   6.12  \\\hline
\end{tabular}

\caption{Stage-wise ablation of GeAR on 1000 samples. We report VEND, DDI, and NIQE for the output w/ or w/o Appearance Restitution.}

\label{step}}
\vspace{-2em}
\end{table}
\noindent\textbf{Stage-wise Effect of Appearance Restitution.}
We  conduct a stage-wise ablation on 1000 samples to examine the role of the two stages in GeAR. Specifically, we compare the grounded output without Appearance Restitution and the final reconstruction with Appearance Restitution, using both quantitative metrics in Table~\ref{step} and qualitative comparisons in Figure~\ref{fig:v-a}. As shown in Table~\ref{step}, adding Appearance Restitution improves both VEND and DDI, indicating that the second stage preserves and further enhances geometric plausibility after Geometry Grounding. This behavior is also visually consistent with Figure~\ref{fig:v-a}: the grounded result exhibits more stable structure but loses part of the original painterly texture, while Appearance Restitution recovers texture, decorative patterns, and stylistic details on top of the grounded reconstruction. As expected, NIQE increases after Appearance Restitution, since this stage intentionally restores artwork-specific textures and stylistic patterns that depart from natural-image statistics.

\begin{table}[!ht]
\setlength{\tabcolsep}{2.85mm}{
\centering
\begin{tabular}{l|c c c}
\hline
\multicolumn{1}{l|}{Prompt Style} 
& \multicolumn{1}{c}{VEND $\uparrow$} 
& \multicolumn{1}{c}{DDI$_{avg}$ $\uparrow$} 
& \multicolumn{1}{c}{NIQE$_{avg}$ $\downarrow$} \\ \hline
\textit{Sculpture}             & \textbf{85.58} & \textbf{11.05} & \textbf{4.79} \\
\textit{Pencil sketch}         & 83.86 & 2.46 &   7.25  \\
\textit{Woodcut}               & 80.46 & 2.95 &   6.79   \\\hline
\end{tabular}

\caption{Results of different prompt styles on geometry-oriented metrics, with NIQE reported as a supplementary diagnostic of natural-image compatibility. }

\label{prompt}}
\vspace{-2em}
\end{table}
\noindent\textbf{Effect of Different Prompt Styles.}
We examine how the \textit{Sculpture} prompt influences reconstruction performance compared to other styles. As shown in Table~\ref{prompt}, the sculpture-style prompt significantly improves both VEND and DDI scores, indicating better 3D structure recovery and more accurate surface details. The physically consistent lighting with clear highlight and shadow transitions provides strong depth cues for the model. In contrast, prompts like \textit{Pencil sketch} and \textit{Woodcut} lack continuous tonal transitions, leading to incomplete shading and reduced geometric fidelity. Furthermore, the sculpture prompt achieves the lowest NIQE, indicating that it better aligns the grounded intermediate with the natural-image statistics assumed by pretrained monocular 3D reconstruction models.

\begin{table}[!ht]
\setlength{\tabcolsep}{3.15mm}{
\centering
\begin{tabular}{c|c c c}
\hline
\multicolumn{1}{l|}{Range of $\alpha$} 
& \multicolumn{1}{c}{VEND $\uparrow$} 
& \multicolumn{1}{c}{DDI$_{avg}$ $\uparrow$} 
& \multicolumn{1}{c}{NIQE$_{avg}$ $\downarrow$} \\ \hline
Fixed         & \textbf{86.01}           & 7.45             & 5.09                 \\
0.5-1.2            & 85.35                    & 10.35            & \textbf{4.79}        \\
0.4-1.5            & 85.29                    & \textbf{11.05}   & \textbf{4.79}        \\
0.3-1.8            & 85.27                    & 11.04            & \textbf{4.79}        \\ \hline
\end{tabular}

\caption{Performance Comparison of Fixed vs. Adaptive $\alpha_{\text{ms}}$ Fusion Coefficient.}
\label{alpha_ad}}
\vspace{-2em}
\end{table}

\noindent\textbf{\bfseries\boldmath $\alpha_{\text{ms}}$ V.S. $\alpha_{\text{adaptive}}$.}
We compare a fixed $\alpha_{\text{ms}}$ with an adaptive $\alpha_{\text{ms}}$ scheme. The fixed setting uses $\alpha_{\text{ms}}=0.8$, while the adaptive scheme is evaluated with $[\alpha_{\min}, \alpha_{\max}] = [0.5, 1.2]$, $[0.4, 1.5]$, and $[0.3, 1.8]$. As shown in Table~\ref{alpha_ad}, the fixed $\alpha_{\text{ms}}=0.8$ achieves a slightly higher VEND, likely because stronger relighting contrast provides more pronounced shading cues for volumetric recovery. However, this more aggressive enhancement also leads to overexposure, which weakens depth-detail preservation and reduces compatibility with the natural-image statistics assumed by pretrained reconstruction models, as reflected by the lower DDI and higher NIQE. 
In contrast, the adaptive scheme yields a better trade-off between structural extent and detail preservation. Among the tested ranges, $[\alpha_{\min}, \alpha_{\max}] = [0.4, 1.5]$ gives the best overall balance, achieving the highest DDI and the lowest NIQE while maintaining competitive VEND. Overall, these results suggest that adaptive $\alpha_{\text{ms}}$ provides a more robust setting for Geometry Grounding across diverse paintings.

\begin{table}[h]
\vspace{-0.5em}
\setlength{\tabcolsep}{1.35mm}{
\centering
\begin{tabular}{cccc}  
\hline
\multicolumn{1}{c!}{Method} 
& \multicolumn{1}{c!}{VEND $\uparrow$} 
& \multicolumn{1}{c!}{DDI$_{avg}$ $\uparrow$} 
& \multicolumn{1}{c}{NIQE$_{avg}$ $\downarrow$} \\ \hline
TRELLIS                      & 77.64 & 7.37 & 5.53         \\
TRELLIS  $+$ HE                    & 75.86 & 3.89 & 5.07         \\
TRELLIS $+$ Retinex                     & 75.40 & 3.24 & 5.39        \\
\rowcolor{blue!5}TRELLIS $+$ Ge & \textbf{85.29} & \textbf{11.05} & \textbf{4.79}     \\ \hline
\end{tabular}

\caption{Comparison of geometry grounding with generic image enhancement baselines. Ge: Geometry Grounding.}
\label{tre}}
\vspace{-2em}
\end{table}

\noindent\textbf{Geometry Grounding vs. Generic Image Enhancement.}
We further compare Geometry Grounding with two generic image enhancement baselines, Histogram Equalization (HE) and Retinex, both integrated into the same TRELLIS reconstruction pipeline (Table~\ref{tre}). Although HE and Retinex slightly reduce NIQE, they substantially degrade VEND and DDI, indicating that generic enhancement alone does not improve geometric recoverability from classical paintings. By contrast, Geometry Grounding consistently improves all three metrics, showing that the gains in Table~\ref{sota-search} do not come from arbitrary preprocessing, but from constructing a more geometry-compatible intermediate representation for monocular 3D reconstruction.

\section{Conclusion}
We introduce \textbf{CP3D}, a new task that aims to reconstruct 3D scenes from classical paintings while preserving fidelity to the source artwork. To address the geometric ambiguity and appearance loss inherent in this setting, we propose \textbf{GeAR}, a training-free two-stage framework that combines Geometry Grounding with Appearance Restitution. We further present \textbf{HeriArch}, a benchmark for systematic evaluation of classical painting-to-3D reconstruction. Through extensive experiments, we show that GeAR produces more stable geometry and more faithful final reconstructions than strong monocular 3D baselines. These results highlight the importance of separating geometric recoverability from appearance restoration in CP3D.

\bibliographystyle{ACM-Reference-Format}
\bibliography{sample-base}

@String(AAAI = {AAAI})

@String(VR   = {Vis. Res.})

@article{songsheng2025re,
  title={Re-examining the Meaning of Flatness Based on the Theory Evolution of Medium Specificity in Painting},
  author={SongSheng, Huang and Tahir, Azian and Ismail, Issarezal},
  journal   = {Asian Journal of Research in Education and Social Sciences},
  volume    = {7},
  number    = {1},
  pages     = {93--104},
  year      = {2025},
  year={2025}
}

@inproceedings{small2019circling,
  title={Circling round Vitruvius, linear perspective, and the design of Roman wall painting},
  author={Small, Jocelyn Penny},
  booktitle={Arts},
  volume={8},
  number={3},
  pages={118},
  year={2019},
  organization={MDPI}
}

@inproceedings{springstein2024visual,
  title={Visual narratives: Large-scale hierarchical classification of art-historical images},
  author={Springstein, Matthias and Schneider, Stefanie and Rahnama, Javad and Stalter, Julian and Kristen, Maximilian and M{\"u}ller-Budack, Eric and Ewerth, Ralph},
  booktitle={Proceedings of the IEEE/CVF Winter Conference on Applications of Computer Vision},
  pages={7220--7230},
  year={2024}
}

@incollection{dahn2024fate,
  title={Fate of the Minotaur: A scalable location based VR experience},
  author={Dahn, Andreas and Plichta, Leszek and Spielmann, Simon and Sch{\"a}fer, Eduard and Bl{\"o}nnigen, Justus},
  booktitle={ACM SIGGRAPH 2024 Immersive Pavilion},
  pages={1--2},
  year={2024}
}

@inproceedings{zhao2025reviving,
  title={Reviving Mural Art through Generative AI: A Comparative Study of AI-Generated and Hand-Crafted Recreations},
  author={Zhao, Shuo and Huang, Yifei and He, Xiaoyang and Tong, Xin and Li, Xin and Wu, Dan},
  booktitle={Proceedings of the 2025 CHI Conference on Human Factors in Computing Systems},
  pages={1--20},
  year={2025}
}

@inproceedings{huang2025spar3d,
  title={Spar3d: Stable point-aware reconstruction of 3d objects from single images},
  author={Huang, Zixuan and Boss, Mark and Vasishta, Aaryaman and Rehg, James M and Jampani, Varun},
  booktitle={Proceedings of the Computer Vision and Pattern Recognition Conference},
  pages={16860--16870},
  year={2025}
}

@inproceedings{chen2024recon3d,
  title={Recon3D: High Quality 3D Reconstruction from a Single Image Using Generated Back-View Explicit Priors},
  author={Chen, Ruiyang and Yin, Mohan and Shen, Jiawei and Ma, Wei},
  booktitle={Proceedings of the IEEE/CVF conference on computer vision and pattern recognition},
  pages={2802--2811},
  year={2024}
}

@article{mildenhall2021nerf,
  title={Nerf: Representing scenes as neural radiance fields for view synthesis},
  author={Mildenhall, Ben and Srinivasan, Pratul P and Tancik, Matthew and Barron, Jonathan T and Ramamoorthi, Ravi and Ng, Ren},
  journal={Communications of the ACM},
  volume={65},
  number={1},
  pages={99--106},
  year={2021},
  publisher={ACM New York, NY, USA}
}

@inproceedings{niemeyer2022regnerf,
  title={Regnerf: Regularizing neural radiance fields for view synthesis from sparse inputs},
  author={Niemeyer, Michael and Barron, Jonathan T and Mildenhall, Ben and Sajjadi, Mehdi SM and Geiger, Andreas and Radwan, Noha},
  booktitle={Proceedings of the IEEE/CVF conference on computer vision and pattern recognition},
  pages={5480--5490},
  year={2022}
}

@inproceedings{somraj2023vip,
  title={Vip-nerf: Visibility prior for sparse input neural radiance fields},
  author={Somraj, Nagabhushan and Soundararajan, Rajiv},
  booktitle={ACM SIGGRAPH 2023 conference proceedings},
  pages={1--11},
  year={2023}
}

@inproceedings{yang2023freenerf,
  title={Freenerf: Improving few-shot neural rendering with free frequency regularization},
  author={Yang, Jiawei and Pavone, Marco and Wang, Yue},
  booktitle={Proceedings of the IEEE/CVF conference on computer vision and pattern recognition},
  pages={8254--8263},
  year={2023}
}

@article{kerbl20233d,
  title={3D Gaussian splatting for real-time radiance field rendering.},
  author={Kerbl, Bernhard and Kopanas, Georgios and Leimk{\"u}hler, Thomas and Drettakis, George},
  journal={ACM Trans. Graph.},
  volume={42},
  number={4},
  pages={139--1},
  year={2023}
}

@inproceedings{charatan2024pixelsplat,
  title={pixelsplat: 3d gaussian splats from image pairs for scalable generalizable 3d reconstruction},
  author={Charatan, David and Li, Sizhe Lester and Tagliasacchi, Andrea and Sitzmann, Vincent},
  booktitle={Proceedings of the IEEE/CVF conference on computer vision and pattern recognition},
  pages={19457--19467},
  year={2024}
}

@article{fan2024instantsplat,
  title={Instantsplat: Unbounded sparse-view pose-free gaussian splatting in 40 seconds},
  author={Fan, Zhiwen and Cong, Wenyan and Wen, Kairun and Wang, Kevin and Zhang, Jian and Ding, Xinghao and Xu, Danfei and Ivanovic, Boris and Pavone, Marco and Pavlakos, Georgios and others},
  journal={arXiv preprint arXiv:2403.20309},
  volume={2},
  number={3},
  pages={4},
  year={2024}
}

@inproceedings{szymanowicz2024splatter,
  title={Splatter image: Ultra-fast single-view 3d reconstruction},
  author={Szymanowicz, Stanislaw and Rupprecht, Chrisitian and Vedaldi, Andrea},
  booktitle={Proceedings of the IEEE/CVF conference on computer vision and pattern recognition},
  pages={10208--10217},
  year={2024}
}

@inproceedings{szymanowicz2025flash3d,
  title={Flash3d: Feed-forward generalisable 3d scene reconstruction from a single image},
  author={Szymanowicz, Stanislaw and Insafutdinov, Eldar and Zheng, Chuanxia and Campbell, Dylan and Henriques, Joao F and Rupprecht, Christian and Vedaldi, Andrea},
  booktitle={2025 International Conference on 3D Vision (3DV)},
  pages={670--681},
  year={2025},
  organization={IEEE}
}

@inproceedings{shue20233d,
  title={3d neural field generation using triplane diffusion},
  author={Shue, J Ryan and Chan, Eric Ryan and Po, Ryan and Ankner, Zachary and Wu, Jiajun and Wetzstein, Gordon},
  booktitle={Proceedings of the IEEE/CVF Conference on Computer Vision and Pattern Recognition},
  pages={20875--20886},
  year={2023}
}

@article{shen2025gamba,
  title={Gamba: Marry Gaussian Splatting With Mamba for Single-View 3D Reconstruction},
  author={Shen, Qiuhong and Wu, Zike and Yi, Xuanyu and Zhou, Pan and Zhang, Hanwang and Yan, Shuicheng and Wang, Xinchao},
  journal={IEEE Transactions on Pattern Analysis and Machine Intelligence},
  year={2025},
  volume={},
  number={},
  pages={1-14},
  publisher={IEEE}
}

@inproceedings{ho2024sith,
  title={Sith: Single-view textured human reconstruction with image-conditioned diffusion},
  author={Ho, I and Song, Jie and Hilliges, Otmar and others},
  booktitle={Proceedings of the IEEE/CVF Conference on Computer Vision and Pattern Recognition},
  pages={538--549},
  year={2024}
}

@inproceedings{wang2025videoscene,
  title={VideoScene: Distilling video diffusion model to generate 3D scenes in one step},
  author={Wang, Hanyang and Liu, Fangfu and Chi, Jiawei and Duan, Yueqi},
  booktitle={Proceedings of the IEEE/CVF Conference on Computer Vision and Pattern Recognition},
  pages={16475--16485},
  year={2025}
}

@inproceedings{sun2025drive,
  title={Drive: Diffusion-based rigging empowers generation of versatile and expressive characters},
  author={Sun, Mingze and Chen, Junhao and Dong, Junting and Chen, Yurun and Jiang, Xinyu and Mao, Shiwei and Jiang, Puhua and Wang, Jingbo and Dai, Bo and Huang, Ruqi},
  booktitle={Proceedings of the Computer Vision and Pattern Recognition Conference},
  pages={21170--21180},
  year={2025}
}

@inproceedings{anciukevivcius2023renderdiffusion,
  title={Renderdiffusion: Image diffusion for 3d reconstruction, inpainting and generation},
  author={Anciukevi{\v{c}}ius, Titas and Xu, Zexiang and Fisher, Matthew and Henderson, Paul and Bilen, Hakan and Mitra, Niloy J and Guerrero, Paul},
  booktitle={Proceedings of the IEEE/CVF conference on computer vision and pattern recognition},
  pages={12608--12618},
  year={2023}
}

@inproceedings{lan2024ln3diff,
  title={Ln3diff: Scalable latent neural fields diffusion for speedy 3d generation},
  author={Lan, Yushi and Hong, Fangzhou and Yang, Shuai and Zhou, Shangchen and Meng, Xuyi and Dai, Bo and Pan, Xingang and Loy, Chen Change},
  booktitle={European Conference on Computer Vision},
  pages={112--130},
  year={2024},
  organization={Springer}
}

@inproceedings{muller2023diffrf,
  title={Diffrf: Rendering-guided 3d radiance field diffusion},
  author={M{\"u}ller, Norman and Siddiqui, Yawar and Porzi, Lorenzo and Bulo, Samuel Rota and Kontschieder, Peter and Nie{\ss}ner, Matthias},
  booktitle={Proceedings of the IEEE/CVF Conference on Computer Vision and Pattern Recognition},
  pages={4328--4338},
  year={2023}
}

@article{xu2024comprehensive,
  title={A comprehensive dataset for digital restoration of Dunhuang murals},
  author={Xu, Zishan and Yang, Yuqing and Fang, Qianzhen and Chen, Wei and Xu, Tingting and Liu, Jueting and Wang, Zehua},
  journal={Scientific Data},
  volume={11},
  number={1},
  pages={955},
  year={2024},
  publisher={Nature Publishing Group UK London}
}

@incollection{wang2018dunhuang,
  title={Dunhuang mural restoration using deep learning},
  author={Wang, Han-Lei and Han, Ping-Hsuan and Chen, Yu-Mu and Chen, Kuan-Wen and Lin, XinYi and Lee, Ming-Sui and Hung, Yi-Ping},
  booktitle={SIGGRAPH Asia 2018 technical briefs},
  pages={1--4},
  year={2018}
}

@inproceedings{li2024towards,
  title={Towards automated Chinese ancient character restoration: a diffusion-based method with a new dataset},
  author={Li, Haolong and Du, Chenghao and Jiang, Ziheng and Zhang, Yifan and Ma, Jiawei and Ye, Chen},
  booktitle={Proceedings of the AAAI Conference on Artificial Intelligence},
  volume={38},
  number={4},
  pages={3073--3081},
  year={2024}
}

@inproceedings{xiang2025structured,
  title={Structured 3d latents for scalable and versatile 3d generation},
  author={Xiang, Jianfeng and Lv, Zelong and Xu, Sicheng and Deng, Yu and Wang, Ruicheng and Zhang, Bowen and Chen, Dong and Tong, Xin and Yang, Jiaolong},
  booktitle={Proceedings of the Computer Vision and Pattern Recognition Conference},
  pages={21469--21480},
  year={2025}
}

@inproceedings{tang2024lgm,
  title={Lgm: Large multi-view gaussian model for high-resolution 3d content creation},
  author={Tang, Jiaxiang and Chen, Zhaoxi and Chen, Xiaokang and Wang, Tengfei and Zeng, Gang and Liu, Ziwei},
  booktitle={European Conference on Computer Vision},
  pages={1--18},
  year={2024},
  organization={Springer}
}

@article{muralnet2022,
  title={Line Drawing Guided Progressive Inpainting of Mural Damages},
  author={Luxi Li and Qin Zou and Fan Zhang and Hongkai Yu and Long Chen and Chengfang Song and Xianfeng Huang and Xiaoguang Wang},
  journal={ArXiv 2211.06649},
  pages={1--12},
  year={2022},
}

@inproceedings{li2025generation,
  title={From generation to judgment: Opportunities and challenges of llm-as-a-judge},
  author={Li, Dawei and Jiang, Bohan and Huang, Liangjie and Beigi, Alimohammad and Zhao, Chengshuai and Tan, Zhen and Bhattacharjee, Amrita and Jiang, Yuxuan and Chen, Canyu and Wu, Tianhao and others},
  booktitle={Proceedings of the 2025 Conference on Empirical Methods in Natural Language Processing},
  pages={2757--2791},
  year={2025}
}

@article{batifol2025flux,
  title={Flux. 1 kontext: Flow matching for in-context image generation and editing in latent space},
  author={Batifol, Stephen and Blattmann, Andreas and Boesel, Frederic and Consul, Saksham and Diagne, Cyril and Dockhorn, Tim and English, Jack and English, Zion and Esser, Patrick and Kulal, Sumith and others},
  journal={arXiv e-prints},
  pages={arXiv--2506},
  year={2025}
}

@inproceedings{chen2024gaussianeditor,
  title={Gaussianeditor: Swift and controllable 3d editing with gaussian splatting},
  author={Chen, Yiwen and Chen, Zilong and Zhang, Chi and Wang, Feng and Yang, Xiaofeng and Wang, Yikai and Cai, Zhongang and Yang, Lei and Liu, Huaping and Lin, Guosheng},
  booktitle={Proceedings of the IEEE/CVF conference on computer vision and pattern recognition},
  pages={21476--21485},
  year={2024}
}

%%
%% If your work has an appendix, this is the 

\end{document}